\documentstyle[12pt]{article}
\newcommand{\doublespacing}{\let\CS=\@currsize\renewcommand{\baselinesstrech}
{2.0}\tiny\CS}

\begin{document}
\setlength{\baselineskip}{18.5pt}

\centerline{\LARGE{\bf  Dark Matter and Supersolidity  }}

\vspace{0.9cm}

\centerline{\large{\bf  Sisir Roy \&
Malabika Roy}}

\begin{center}
{\bf Physics \& Applied Mathematics Unit \\
Indian Statistical Institute \\
Calcutta-700 035, India \\
e-mail : sisir@isical.ac.in }
\end{center}

\vspace{1cm}

\begin{center}
{\bf Abstract}
\end{center}
The recet experiments on supersolidity sheds new light on the issue of Dark matter and the missing mass.
The origin of collisionless cold dark matter has been traced back to its origin to the supersolid model of quantum vacuum.
\noindent
\vskip5pt
\newpage

\section{\bf Introduction} 
The issue of Dark matter raised lot of controversy over the past decade. The universe is supposed to be full of "dark matter"
that influences the evolution of the universe gravitationally but not observed directly.
The Cambridge team put a limit on how it is packed in space. Their observations claimed that this dark matter makes ${85}$
percent of the universe. Very recently, the European space agency found the possible distribution of the dark matter in the universe.
Dark matter are divided into two broad groups
\begin{enumerate}
\item Hot Dark matter(HDM)
\item Cold dark matter(CDM)
\end{enumerate}
Particles of zero or near-zero mass are main constituents of Hot Dark matter while Cold Dark Matter are mainly composed of massive particles moving at sub-relativistic velocities. The HDM palys different role than CDM in structure formation because the high velocities of HDM wipes out the atructure at small scale.
The recent experiments$^{1,2}$ at low tempearture indicates a new state of matter popualrly known as supersolid.  Any individual wave-packet associated with a single atom increases in size as it is being cooled especially around absolute zero where as at higher temperature atoms are locked in a grid. Thus atoms loose their individuaklity at very low temperature and become one giant atom.
The results attarcted lot of attention because the characteristics of solid and liquid is describable only within the weired world of quantum mechanics. It is wellknown that a solid can not flow so it can not exhibit some of the properties of liquid like superfluid.
However, when the above superlfuid is put under pressure slightly above absolute zero, a portion of it becomes supersolid and looses its friction completely. Here, the crystal suddenly became lighter but the missing mass reappeared when it is slightly heated.
Several authors$^{3}$ pointed out that the vacuum can be thought of as type of superfluid and others considered it as with little rsistance or non-zero conductivity$^{4,5}$. They tried to explain redshift using the concept of tired light. 
Here, we propose that the vacuum is composed  of  supersolid like matter which can be thgouht of as collisonalless cold dark matter. This sheds new light on the origin of  missing mass. 
We shall discuss the dark matter and the issue of missing mass in section II.
Possible implications are discussed in section III. 
\noindent 
\vskip5pt
\section{\bf Hot and Cold Dark Matter }
Astronomer Zwicky$^{6,7}$ cointed the term Dark Matter. In 1933  during his meausrement of individual velocities of a large group of galaxies , the galaxies are found
to be moving with so rapidly that they should come apart. The visible mass of the galaxies in the cluster can not produce enough gravitational force to confine the galaxies within the cluster. He speculated perhaps the missing mass do not give off or reflect enough light to see them.
The cosmologists working with Big Bang hypothesis faced an entirely diffrent problem.They found that there is not enough gravitational energy in the cosmos so as to compatible with General Theory of Relativity. They called it as Dark energy. If one beleives that gravitational force is the only force in the cosmos,then $^8$
of the universe had to be filled with missing entities. 
Dark matter are broadly classified in to two groups :Hot Dark Matter and Cold Dark Matter.
The fast moving particles like neutrinos,tachyons are supposed to be constituent of HDM and massive particles are for CDM.
Ostreiker and Steinhardt$^{9}$ made a list for several type of dark matter :
\begin{enumerate}
\item Cold Collisionless Dark Matter (CCDM)
\item Strongly Self interacting Dark Matter
\item Warm Dark Matter
\item Repulsive Dark Matter
\item Self annihilating dark matter
\item Fuzzy dark matter
\end{enumerate}
The predictions of cold dark matter have been analyzed with respect to the present observational evidences. Moreover, it is not at all clear about the
nature of cold dark matter particles. Lambda-Cold Dark Matter has also been discussed as concordanc model of big bang cosmology. However, it says nothing about the 
physical origin of dark matter.
Here, we consider the quantum vacuum as consisiting of supersolid like matter. Chan and Kim$^{2}$ observed that when a particular isotope of helium gas has frozen into a crystal at a fraction of a degree above absolute zero, part of it exihibits a property with no friction as observed in superfluids.
It seems some mass is missing and by reheating it inreases the resistance and mass reappaers. 
This fluid is frictionless ie.  like collisionless matter not observed directly. So the promordial supersolid is like collisionless dark matter.
It is possible to trace back to the physicalorigin of this type of cold dark matter.

\section{\bf Possible Implications }
\noindent
\vskip5pt
The above analysis sheds light on the physical origin as well as the possibility of observing dark matter.
Moreover, the distribution of dark matter can be studied within this framework.

\noindent
\vskip5pt
\newpage

\section {References}
\noindent
\vskip5pt
\begin{enumerate}

\item E. Kim and M.H.W.Chan ,(2004), Nature {\bf 427},225. 

\item J. Day and J.Beamish,(2007), Nature {\bf 450}, 853.

\item  A.Rothwarf, (1998)Physics Essays {\bf 11}, 444.

\item B.Lehnert and S.Roy(1998), {\it Extended Electromagnetic Theory}, World Scientific Publishers, Singapore.

\item S.Roy,G.Kar and M.Roy(1996)Int.Journ. Theo. Phys. {\bf 35}, 579.

\item F. Zwicky (1933) Helvetica Physica Axta {\bf 6}, 110.

\item F. Zwicky (1937) Astrophysical Journal {\bf 86}, 217.

\item S. Perlmutter(2003) Physics Today, April, 53.

\item J. P. Ostriker and P. Steinhardt(2003),Science {\bf 300}, 1909.
\end{enumerate}

\end{document}